# Near-Field Coupling of Polypropylene Dielectric Waveguide Routed Near PCB Board at Terahertz Frequencies

Wenbo Liu[a], Jiabiao Zhao[a], Kefeng Huang[a], Peian Li[a], Boquan Xu[a], Yang Cao[b], Weidong Hu[a] and Jianjun Ma[a,*]

[a]*School of Integrated Circuits and Electronic, Beijing Institute of Technology, and the State Key Laboratory of Environment Characteristics and Effects for Near-space, Beijing 100081, China.*

[b]*School of Electronic and Information Engineering, Hebei University of Technology, Advanced Laser Technology Research Center, Tianjin 300130, China.*

***Abstract*—Terahertz (THz) dielectric waveguides are promising physical channels for short-reach interconnects, but their air-clad guided fields may interact with nearby printed-circuit-board (PCB) structures in compact packages. In this work, we experimentally and numerically investigate PCB-proximity-induced excess transmission loss in 3D-printed polypropylene rectangular dielectric waveguides over 220-325 GHz. Continuous-wave transmission measurements are performed for bare FR4, continuous waveguide-facing copper, and periodic copper-trace PCB configurations under controlled clearance and alignment conditions. The results show that direct contact with bare FR4 can induce a frequency-selective high-loss band, which is attributed to phase-matched leakage from the guided waveguide mode into a substrate-supported leaky branch. This finding highlights PCB proximity as a critical layout factor and provides practical guidance for clearance control and metallization design in compact THz dielectric-waveguide packages.**

## 1. Introduction

The continued scaling of data-centric computing, heterogeneous integration, and chiplet-based architectures has placed unprecedented demands on short-reach interconnects in terms of aggregate throughput, latency, energy per bit, and packaging density [1]. Because data movement now dominates over computation in the energy budget of high-performance systems, the migration of interconnect technology toward the millimeter-wave and terahertz (THz) regimes offers a credible route past the bandwidth ceilings imposed by conventional electrical signaling [2, 3]. The interest in THz interconnects extends beyond conventional chip-to-chip and board-level routing to a broader class of compact links, including antenna-to-transceiver feeds, front-end-to-back-end routing within System-in-Package modules [2-4], and short-reach connections among distributed sensing, computing, and communication elements in next-generation electronic platforms. In this frequency range, however, metallic transmission media become increasingly unattractive: conductor loss scales unfavorably with frequency, dielectric-loading penalties grow, parasitic coupling among neighboring traces strengthens, and packaging-induced impedance discontinuities are difficult to control on the relevant length scales [3, 5, 6]. The cumulative effect is degraded signal integrity and a hard limit on scalable integration, motivating the search for alternative guided-wave platforms suited to THz operation.

Against this backdrop, dielectric waveguides have attracted growing attention as a guided medium for the THz regime. By confining electromagnetic energy primarily within a low-loss dielectric core and the surrounding air rather than within a conductor [7], dielectric waveguides largely sidestep conductor-dominated dissipation while retaining broadband, low-dispersion propagation. Polymer-based platforms in particular are compatible with low-cost additive manufacturing such as fused-deposition-modeling 3D printing and extrusion [8, 9], can be interfaced directly with electronic THz front-ends through rectangular-to-dielectric waveguide transition [10, 11], and avoid the optoelectronic conversion overhead intrinsic to optical-fiber interconnects [12]. Gigabit-class data transmission and functional routing through dielectric-waveguide-based THz links have been experimentally demonstrated [13], confirming their practical relevance as short-reach, high-capacity interconnect media. These characteristics position THz dielectric waveguides as an enabling technology for compact, highly integrated electronic and electronic-photonic systems.

TABLE 1
REPRESENTATIVE STUDIES ON TERAHERTZ DIELECTRIC WAVEGUIDES

| Topic | Refs. | Typical Scope | Contributions |
|---|---|---|---|
| Waveguide structure engineering | [14-17, 56, 57] | Porous fibers, suspended-core fibers, Si photonic-crystal slab, effective-medium cladding | Improve confinement and reduce intrinsic guiding loss |
| Material platform and characterization | [18-21] | TOPAS, Zeonex, and related low-loss dielectric platforms | Establish low-loss material basis for THz guidance |
| Propagation loss, dispersion, and bend | [14, 22, 23] | Straight-waveguide attenuation, modal purity, dispersion, bending loss | Target low-loss, low-dispersion, and practical routing |
| Integrated routing and interconnects | [9, 10, 24, 25] | On-chip routing, topological transport, waveguide interfaces, high-speed links | Demonstrate interconnect potential at system level |
| Evanescent-field interaction | [26-28] | Sensing and spectroscopy via surrounding-medium perturbation | Confirms that guided modes are sensitive to the external environment |

Over the past decade, extensive studies summarized in Table 1 have established the structural, material, and link-level foundations of THz dielectric waveguides. Waveguide-geometry engineering has spanned porous fibers [14], suspended-core designs [15], silicon photonic-crystal slabs [16], and effective-medium-cladded structures [17]; low-loss material development has covered TOPAS, Zeonex, polypropylene (PP), and related polymers [18-21]; detailed characterization of propagation loss, modal purity, dispersion, and bend-induced attenuation has informed practical routing geometries [14, 22, 23]; and integrated-routing and link-level demonstrations have established the viability of dielectric-waveguide interconnects at the system level [9, 10, 24, 25]. In a complementary direction, evanescent-field interactions have been deliberately exploited in sensing and spectroscopy, where the sensitivity of the guided mode to its surrounding medium becomes the operating principle rather than a parasitic effect [26-28].

Despite this progress, the behavior of bare dielectric waveguides routed in close proximity to realistic PCB substrates remains insufficiently quantified [29, 30]. In practical packages, FR4, solder mask, copper traces, ground planes, and component metallization can lie within the evanescent interaction region of an air-clad THz waveguide. For the $1\times1$ mm² PP waveguide studied here, the calculated air-side 1/e amplitude decay length is on the order of 150-250 μm across 220-325 GHz, so sub-millimetre clearance variations and weak residual tails can substantially alter leakage into nearby materials. Prior sensing work deliberately exploits this environmental sensitivity; in package-level interconnects, however, the same interaction becomes a parasitic loss channel. The present work therefore focuses not on the existence of evanescent coupling, but on its magnitude, spectral selectivity, distance dependence, and dependence on PCB metallization geometry. We quantify the excess loss induced by bare FR4, continuous copper, and periodic copper traces; identify a phase-matching condition associated with an FR4 slab branch; and translate the results into practical routing rules for compact THz dielectric-waveguide links.

## 2. Waveguide design and measurement setup

Polypropylene (PP) is selected as the guiding material for its low THz dielectric loss, weak dispersion ($n\approx1.51$, weakly frequency-dependent across 220-325 GHz [31-33]), and FDM-printability. Waveguides are printed on an Ultimaker 2+ Connect (XYZ resolution $12.5\times12.5\times5$ μm$^3$, layer 200/20 μm) in an air-cladded geometry that reflects bare-interconnect deployment rather than enclosed-channel designs. The cross-sectional dimension trades off modal content, field confinement, mechanical rigidity, and environmental sensitivity. Sub-millimeter cores such as $0.6\times0.6$ mm$^2$ approach quasi-single-mode operation but enlarge the evanescent extent and reduce flexural stiffness (see Fig. S1, S4, S5 in Supplemental Document).

Larger cores improve confinement at the cost of higher-order mode admission [20, 34-36]. We adopt 1×1 mm$^2$ as a balanced compromise. The fundamental hybrid pair $E_x^{11}/E_y^{11}$ dominates throughout the band with single-lobed transverse profiles (see Fig. S2 in Supplemental Document), higher-order branches are weakly confined and operationally negligible at the upper band edge (see Fig. S3 in Supplemental Document), and gravitational sag is sub-100 μm at the geometry used (see Fig. S4 in Supplemental Document). The waveguide is routed in a U-shape consisting of one 160 mm straight section and two 150 mm straight sections joined by 90° bends of radius R=20 mm, for a total guided path of ≈523 mm; bend loss at this radius is negligible [9].

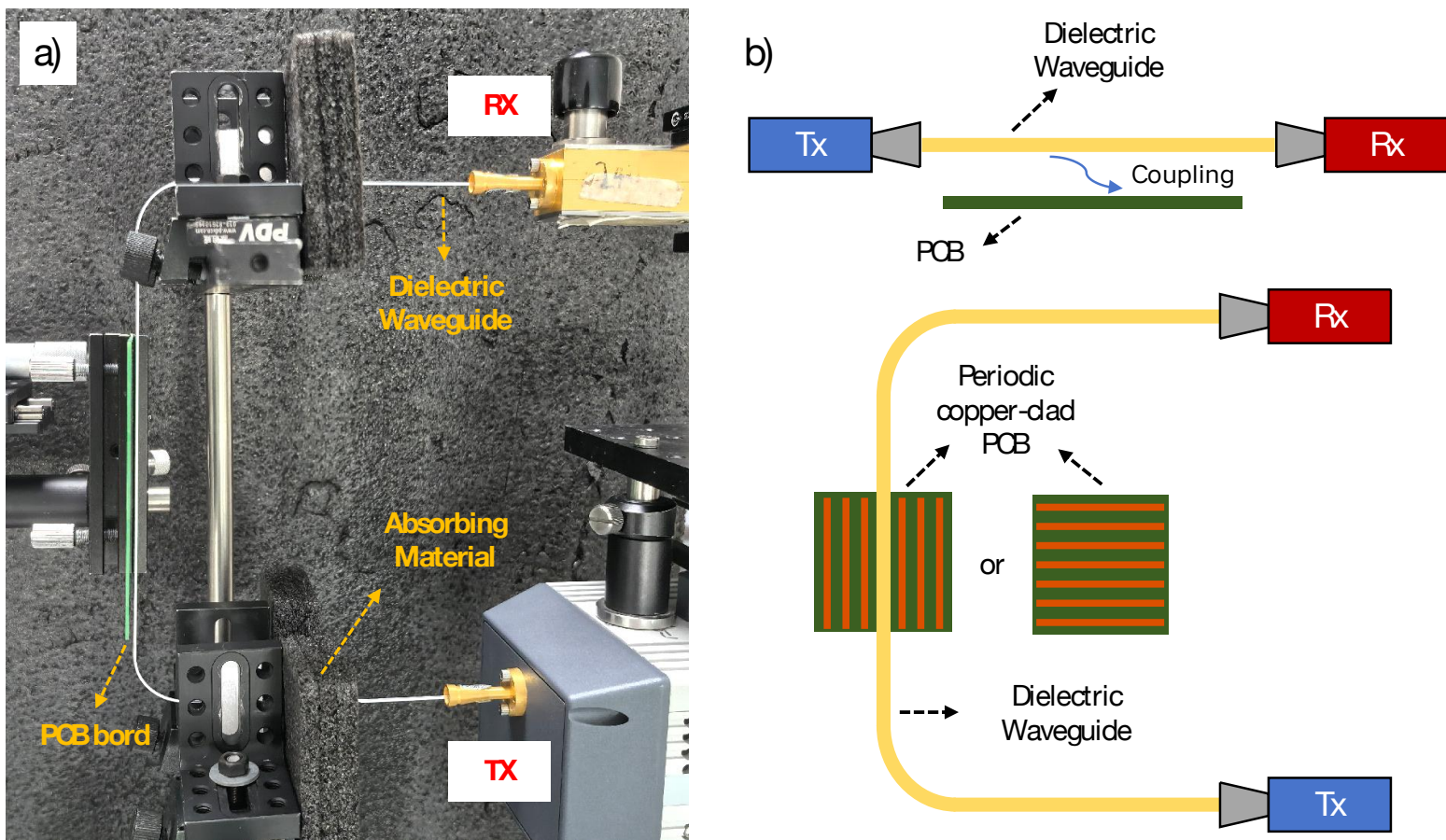

**Fig. 1.** Experimental setup and measurement configurations for the PP dielectric waveguide near PCB structures. (a) Photograph of the waveguide-PCB measurement setup; (b) schematic of the coupling configurations for bare and periodic copper-trace PCBs.

Continuous-wave transmission is measured over 220-325 GHz using a Ceyear 1465D source, an 82406D multiplier chain, and an 89901S horn antenna at the transmit side, and a matched horn coupled to a calibrated 87106B sensor / 71718 power meter at the receiver. Both horns sit on the same side of the optical bench (Fig. 1(a) and (b)). The U-routed waveguide ensures that the received signal is dominated by guided rather than free-space line-of-sight power. The horn-to-waveguide insertion is fixed at ≈20 mm; collinear alignment with the horn boresight is maintained across all measurements by a common mechanical holder, so only the PCB position is varied. Absorber tiles cover all support structures to suppress parasitic scattering [37]. A 10×10 cm$^2$ PCB is placed parallel to one face of the bare waveguide, defining a 100 mm interaction length. A high-resolution translation stage sets the waveguide-PCB clearance (g). PCB samples include bare FR4 (1.06 mm thick, 30 μm solder mask on each face), single-sided full-copper (20 μm Cu beneath the waveguide-facing solder mask), and four periodic-trace PCBs (1 mm trace width; 1, 2, 3, and 4 mm inter-trace gaps; copper coverage of approximately 49%, 33%, 25%, and 20%, respectively). Three trace-waveguide alignments are tested - perpendicular, parallel on-trace, and parallel between-traces. Measurements are performed at 24 °C and 30% relative humidity (RH). Reference and PCB-loaded transmission share the same path and ambient, so atmospheric contribution cancels in the subtraction.

## 3. Transmission behavior near bare and fully metallized PCBs

### 3.1 Excess-loss extraction methodology

For an air-cladded dielectric waveguide, the guided field is not fully confined within the dielectric core but extends into the surrounding region as an evanescent tail. The propagation characteristics are therefore inherently sensitive to nearby material boundaries [38]. When dielectric or metallic objects intrude into this interaction region, the local electromagnetic boundary conditions are modified, perturbing the modal field distribution, opening additional radiation or leakage channels, and altering the effective mode confinement [26, 39]. In compact THz packages, where dielectric waveguides are routed close to PCB substrates and copper traces, this near-field sensitivity becomes a packaging-level design issue rather than a secondary perturbation.

To quantify the near-field-coupling-induced (NFC-induced) transmission degradation, we extract the excess loss as $\Delta P_{\mathrm{loss}}(f, g)=P_{\mathrm{ref}}(f)-P_{\mathrm{PCB}}(f, g)$, where $P_{\mathrm{ref}}(f)$ is the reference transmission measured without the PCB and $P_{\mathrm{PCB}}(f, g)$ is the transmission measured with the PCB at clearance g. Three clearances are investigated with $g = 0$, 0.5, and 1 mm, using a 10×10 $cm^2$ PCB sample under controlled indoor conditions (24 °C, 30% RH). Because reference and PCB-loaded transmission are recorded over the same path under identical ambient conditions, atmospheric attenuation cancels in the subtraction and does not contribute to the extracted excess loss. The reference-subtraction procedure also suppresses common-path effects from horn-to-waveguide insertion mismatch, residual fixture reflections, and slow drift of the source power, isolating the attenuation specifically introduced by the PCB.

### 3.2 PCB stack-up and dielectric characterization

Two PCB configurations with identical overall dimensions are first considered - a bare PCB without metallization, and a single-sided PCB with full copper coverage on the waveguide-facing side. Both samples share an identical FR4 substrate (thickness $t = 1.06$ mm) with 30 μm solder-mask layers on both faces. In the metallized case, a 20 μm copper layer is inserted between the waveguide-facing solder mask and the FR4. From the waveguide-facing surface inward, the layer sequence is therefore solder mask - Cu (if present) - FR4 - solder mask. This pairing isolates two competing effects of dielectric loading by the lossy FR4 and metallic shielding by the copper layer.

The dielectric properties of the PCB stack are characterized using a T-SPEC 800 THz time-domain spectroscopy (THz-TDS) system operating across 0.1-3 THz [40]. The system employs a voice-coil-driven high-speed delay line, providing a 116 ps time-domain scan in 0.1 s and a corresponding spectral resolution of approximately 8 GHz. The complex refractive index is retrieved from the amplitude ratio and phase delay of the Fourier-transformed sample and reference time-domain signals, with the real part n obtained primarily from the phase delay and the imaginary part κ from the frequency-dependent amplitude attenuation, after accounting for sample thickness. Within the 220-325 GHz band, the extracted real part remains nearly constant at $n \approx 2.23\text{-}2.27$, while the imaginary part decreases gradually from $\kappa \approx 0.10$ at 220 GHz to $\kappa \approx 0.05$ at 325 GHz (see Fig. S6 in the Supplemental Document). No sharp absorption resonance appears in this range, which makes the FR4 substrate behaves as a broadband lossy dielectric rather than a resonant absorber. Repeated THz-TDS measurements confirm that the 30 μm solder-mask coating contributes negligibly to the effective dielectric response, so the dominant substrate contribution originates from the FR4 itself. The loss tangent, $\tan \delta = 2n\kappa/(n^2 + \kappa^2)$, decreases from approximately 0.089 near 220 GHz to 0.044 near 325 GHz, consist
ent with previously reported sub-THz characterizations of FR4 [41].

### 3.3 Finite-element simulation configuration

To support the experimental interpretation, two-dimensional frequency-domain finite-element simulations are performed using a parametric sweep over 220-325 GHz with 2 GHz spectral resolution. Input and output ports are defined through boundary-mode analysis to launch and receive the fundamental hybrid mode of the 1×1 $mm^2$ PP waveguide, after which a full-wave field calculation is carried out in the frequency-domain solver. Scattering boundary conditions are applied at the outer simulation boundaries to suppress non-physical reflections, and the maximum mesh size is constrained to one-eighth of the local wavelength to resolve both the guided mode and its evanescent tail. The waveguide-PCB interaction is modeled as a 100 mm parallel coupling section with uniform separation (clearance) g, matching the experimental interaction length; the excess loss is extracted relative to the corresponding no-PCB reference simulation. Material parameters are $n_{\mathrm{PP}} = 1.51$ with negligible loss [31], the THz-TDS-extracted complex permittivity for FR4, and a Drude-model copper conductivity of $\sigma_{\mathrm{Cu}} \approx$ 5.8×107 S/m [42]. The two-dimensional approximation is justified by the longitudinal uniformity of the parallel-coupling geometry. Three-dimensional effects from finite PCB edges and U-bend curvature are not modeled and are addressed separately as residual sources in Section 3.5. The two-dimensional treatment is applied only to the bare-FR4 and full-Cu configurations. There it is a consequence of symmetry - over the 100 mm coupling region the FR4 substrate and the continuous copper layer are translationally invariant along the propagation direction and laterally uniform across the few-hundred-micron extent of the guided mode and its evanescent tail (Section 3.4), so the problem is separable and the reduced-dimensional model reproduces the full three-dimensional physics at a fraction of the cost. The simulation-measurement agreement for both PCB types in Fig. 2 - band centre, width, and the smooth full-Cu response - validates this. Finite PCB-edge scattering and U-bend curvature are the only three-dimensional contributions to these two cases and are treated as residual sources in Section 3.5.

### 3.4 Bare and full-Cu PCB results

Bare FR4, direct contact ($g$ = 0). Fig. 2(a) shows that direct contact with the bare FR4 substrate produces an excess loss that exceeds the measurement dynamic-range limit (30-32 dB measured; FEM predicts ≥ 45 dB) across a pronounced high-loss band centered between 235 and 265 GHz. The repeatability analysis in Fig. S8 confirms that this band coincides with elevated measurement standard deviation (up to ≈ 10 dB), consistent with operation near the noise floor of the receiver chain. The measured excess loss in this region should therefore be interpreted as a lower-bound estimate rather than a precise peak value. The band is used to identify the occurrence of strong substrate-assisted leakage, not to quantify its maximum.

The frequency-localized character of the high-loss band is more consistently explained by a phase-matched leakage condition [43] than by intrinsic FR4 absorption. The substrate κ varies smoothly across the band (Fig. S6), whereas the measured loss exhibits a sharp 25-30 GHz wide peak. This frequency-selective band is consistent with substrate-assisted leakage from the PP-guided mode into lossy FR4 slab-supported fields [44]. With $t$ = 1.06 mm and $\tilde{n}_{\mathrm{FR4}} \approx 2.25 + j\cdot 0.07$, the FR4 slab supports a sequence of leaky transverse modes [45] whose effective indices cross the PP-guided index $n_{\mathrm{eff}} \approx$ 1.28-1.37 (extracted via the effective-index method (EIM) and shown in Fig. S3) within the 255-265 GHz window. The approximate cutoff of the relevant slab branch, $f_{c,m} \approx mc/\left(2t\sqrt{n_{FR4}^{2}-1}\right)$ ≈ 210 GHz for $m$ =3, places the third higher-order branch in the lower part of the measurement band, consistent with the observed peak position. This interpretation is corroborated by prior demonstrations of integrated tunneling between dielectric waveguides and adjacent dielectric slabs at THz frequencies [39], where similar frequency-selective leakage channels appear in the absence of any intrinsic material resonance.

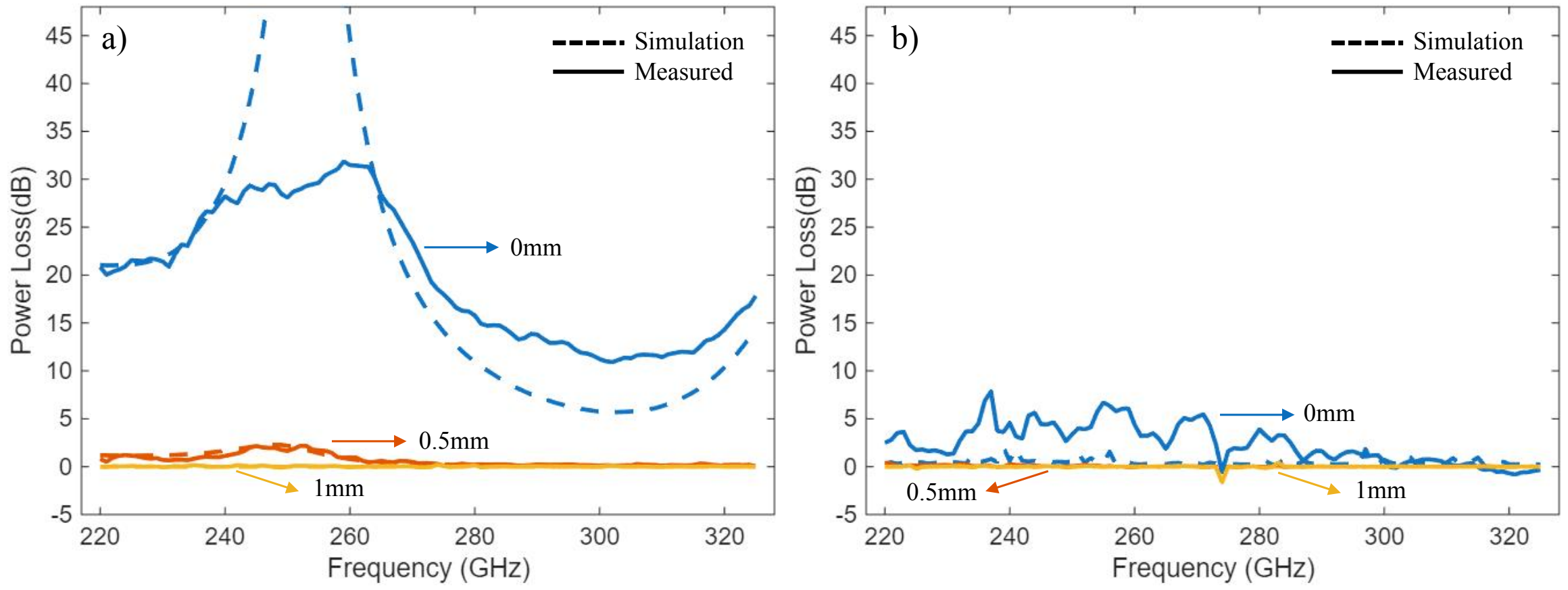

**Fig. 2.** Measured and simulated excess power loss induced by waveguide-PCB near-field coupling: (a) bare PCB without metallization; (b) single-sided PCB with full copper coverage on the waveguide-facing side.

With full-Cu coverage on the waveguide-facing side, Fig. 2(b) shows that the full-Cu PCB produces a substantially lower excess attenuation, peaking at approximately 8 dB and exhibiting a smooth, structureless spectral response across the band. The fundamental change is mechanistic. The continuous copper layer enforces a near-zero tangential E-field at the waveguide-facing surface, blocking the evanescent field from coupling into the lossy FR4 and effectively eliminating the substrate-assisted leakage channel. The remaining ≤ 8 dB residual loss is attributed to finite ohmic loss in the copper layer driven by induced surface currents, with skin depth $\delta_s \approx \sqrt{2/\left(\omega\mu_0\sigma_{Cu}\right)}$ ≈ 0.132 μm at 250 GHz - well below the 20 μm copper thickness, so the layer behaves as an electrically thick conductor. The second contribution may be the modal perturbation from the metal boundary, which slightly modifies the effective index of the guided mode, and the third may be scattering at the finite PCB perimeter. Negative excess-loss values within ± 1 dB of zero in Fig. 2(b) lie within measurement uncertainty and indicate negligible PCB-induced loss rather than physical gain.

To compare the two PCB types compactly, we define a band-averaged shielding factor as

$$\eta_{shield} = 1 - \frac{\langle \Delta P_{loss} \rangle_{full-Cu}}{\langle \Delta P_{loss} \rangle_{bare}} \tag{1}$$

where $\langle\cdot\rangle$ denotes the band average over 220-325 GHz. Using the $\langle \Delta P_{loss} \rangle$ values reported in Table 2 ($\langle \Delta P_{loss} \rangle_{bare}$ = 19.9 dB, $\langle \Delta P_{loss} \rangle_{full-Cu}$ = 2.4 dB), $\eta_{shield} \approx 0.88$, confirming that the continuous metal layer functions primarily as a shielding boundary rather than as a competing loss channel in the tested geometry.

For both PCB types, the excess loss decreases rapidly as $g$ increases from 0 to 0.5 mm and then to 1 mm, approaching noise-floor levels at the largest clearance. This strong distance dependence is the signature of evanescent-field interaction. The field amplitude reaching the substrate decays as $\exp(-\kappa_{air} \cdot g)$, so the leakage power scales as $\exp(-2\kappa_{air} \cdot g)$. With $\kappa_{air} \approx 4.72\times10^3$ m$^{-1}$ at 250 GHz from the EIM, the predicted 1/e decay length is ≈ 212 μm, consistent with the observation that excess loss collapses to within measurement uncertainty by $g$ = 1 mm.

To assess whether PCB proximity also perturbs the modal dispersion relevant to broadband signaling, the differential group-velocity dispersion (GVD) of the guided mode was evaluated relative to the no-PCB reference at $g$ = 0, 0.5, and 1 mm (Fig. S7). The differential GVD remains within ± $1\times10^{-7}$ ps²/m across the band, with only minute fluctuations attributable to discrete-frequency numerical differentiation. We conclude that PCB proximity introduces a negligible perturbation to the second-order frequency dependence of the propagation constant. The degradation observed in Fig. 2 is therefore an attenuation problem, not a dispersion problem - a useful distinction for broadband THz signaling, because it implies that PCB-proximity-induced impairments can be addressed by clearance budgeting and metallization design without introducing pulse-broadening penalties.

### 3.5 Phenomenological model of leakage attenuation

To consolidate the bare-FR4 measurements into a closed-form representation, we develop a phenomenological descriptor that captures the two robust features of the dataset - exponential decay of the excess loss with waveguide-PCB clearance, and finite-band leakage enhancement near the phase-matching frequency - within a framework grounded in evanescent-field theory. The descriptor is not intended as a substitute for full-wave analysis of arbitrary PCB layouts. Rather, it is intended to provide a physically interpretable, four-parameter summary of the dominant loss mechanism over the well-constrained portion of the ($f$, $g$) parameter space, and a documented extrapolation, with caveats, over the region in which the measurement is dynamic-range limited.

The lateral extent of the guided field into the cladding is governed by the propagation constant $\beta(f)$ and the air-side decay constant $\kappa_{air}(f)$, from the effective-index method (EIM) [46, 47], as

$$\beta(f) = k_0 n_{eff}(f) \tag{2}$$

$$\kappa_{air} = \sqrt{\beta^2 - (k_0 n_{air})^2} \tag{3}$$

with $k_0 = 2\pi f / c_0$, $n_{air} \approx 1$ and $f$ is the operating frequency, $c_0$ is the speed of light in vacuum. The field amplitude in the air cladding decays exponentially [48], as

$$E(x, f) \propto \exp\left[-\kappa_{air}(f) x\right] \tag{4}$$

So, at a waveguide-PCB clearance $g$, the field amplitude reaching the substrate is

$$E(g, f) \propto \exp\left[-\kappa_{air}(f) \cdot (g + g_0)\right] \tag{5}$$

where the constant $g_0$ absorbs the residual surface roughness, layer-thickness tolerance, and assembly clearance that prevent the nominal $g$ = 0 condition from corresponding to ideal mechanical contact. With the leakage power scaling as the squared field amplitude, the per-unit-length leakage attenuation coefficient (in Np/m of power) is written as

$$\alpha_{rad}(f, g) = \alpha_0 S(f) \exp\left[-2\kappa_{air}(f)(g + g_0)\right] \tag{6}$$

in which $\alpha_0$ is the peak per-unit-length leakage rate evaluated at $f = f_c$ and $g = -g_0$, and $S(f)$ is a normalized Gaussian spectral envelope with $S(f_c) = 1$, representing the frequency selectivity of the substrate-coupling channel. The normalized spectral envelope is modeled as a Gaussian function of the logarithmic frequency variable $\ln(f/f_c)$, as

$$S(f) = \exp\left[-\frac{\left(\ln(f/f_c)\right)^2}{2\sigma_{\ln f}^2}\right] \tag{7}$$

where $\sigma_{\ln f}$ is the dimensionless standard deviation in the logarithmic frequency domain, rather than a full-width-at-half-maximum (FWHM).

The half-maximum frequencies are obtained from $S(f_\pm) = 1/2$ as $f_\pm = f_c \exp\left(\sigma_{\ln f}\sqrt{2\ln 2}\right)$ giving the equivalent linear-frequency FWHM $\Delta f_{FWHM} = f_+ - f_- = 2f_c \sinh\left(\sigma_{\ln f}\sqrt{2\ln 2}\right)$. Expressed in dB over the experimental interaction length $L_{\text{int}}$ = 0.10 m, the descriptor output is

$$\Delta P_{loss}(f,g) = 4.343 L_{\text{int}}\, \alpha_{rad}(f,g) \tag{8}$$

A nonlinear least-squares fit of Eq. (8) to the measured $\Delta P_{\text{loss}}(f, g)$ heat map of Fig. 3(a) yields the parameter values $\alpha_0 \approx$ 597 Np/m, $g_0 \approx 100$ μm, $f_c \approx 250$ GHz, and $\sigma_{\ln f} \approx 0.0458$ corresponding to a FWHM of about 30 GHz, with 95% confidence intervals reported in the parameter table. Critically, the fit is restricted to data points outside the dynamic-range-limited region (approximately 235-265 GHz at g ≤ 0.3 mm, where the measured response saturates at the receiver noise floor near 30 dB and the FEM result is itself bounded from above by simulation dynamic-range limits at ≈ 45 dB). The model's behavior at peak loss is therefore an extrapolation from the wings of the leakage band into the contact-peak region, not a constrained fit.

The fit achieves RMSE = 3.96 dB and $R^2 = 0.734$ over the constrained region of $(f, g)$ space, with residuals concentrated at the high-frequency band edge (≥ 300 GHz). We interpret these figures as evidence that the model captures the gross $(f, g)$ dependence rather than as a high-fidelity match, and we do not advance any stronger claim. The EIM-predicted air-side decay constants are $\kappa_{\text{air}}$(220 GHz) ≈ 3.96×10$^3$ m$^{-1}$, $\kappa_{\text{air}}$(250 GHz) ≈ 4.72×10$^3$ m$^{-1}$, and $\kappa_{\text{air}}$(325 GHz) ≈ 6.60×10$^3$ m$^{-1}$, corresponding to 1/e amplitude decay lengths of 253, 212, and 152 μm, respectively. These decay lengths are consistent with the observed collapse of the excess loss to within measurement uncertainty by $g = 1$ mm across the full band.

Evaluating the descriptor at the phase-matching frequency under nominal contact, $\Delta P_{\text{loss}}(f_c, g{=}0) \approx 101$ dB, which exceeds the FEM peak (≥ 45 dB) by approximately a factor of two in dB and the measured saturation level (≥ 30 dB) by approximately a factor of three. Since the fit excludes the dynamic-range-limited region from which a constraint on the peak could have been obtained, the descriptor over-predicts in the precise region whose physics is most diagnostic. We therefore recommend that the model be used quantitatively only for $g \geq 0.3$ mm and outside the 235-265 GHz region, and qualitatively elsewhere, with the contact-peak prediction treated as an upper bound.

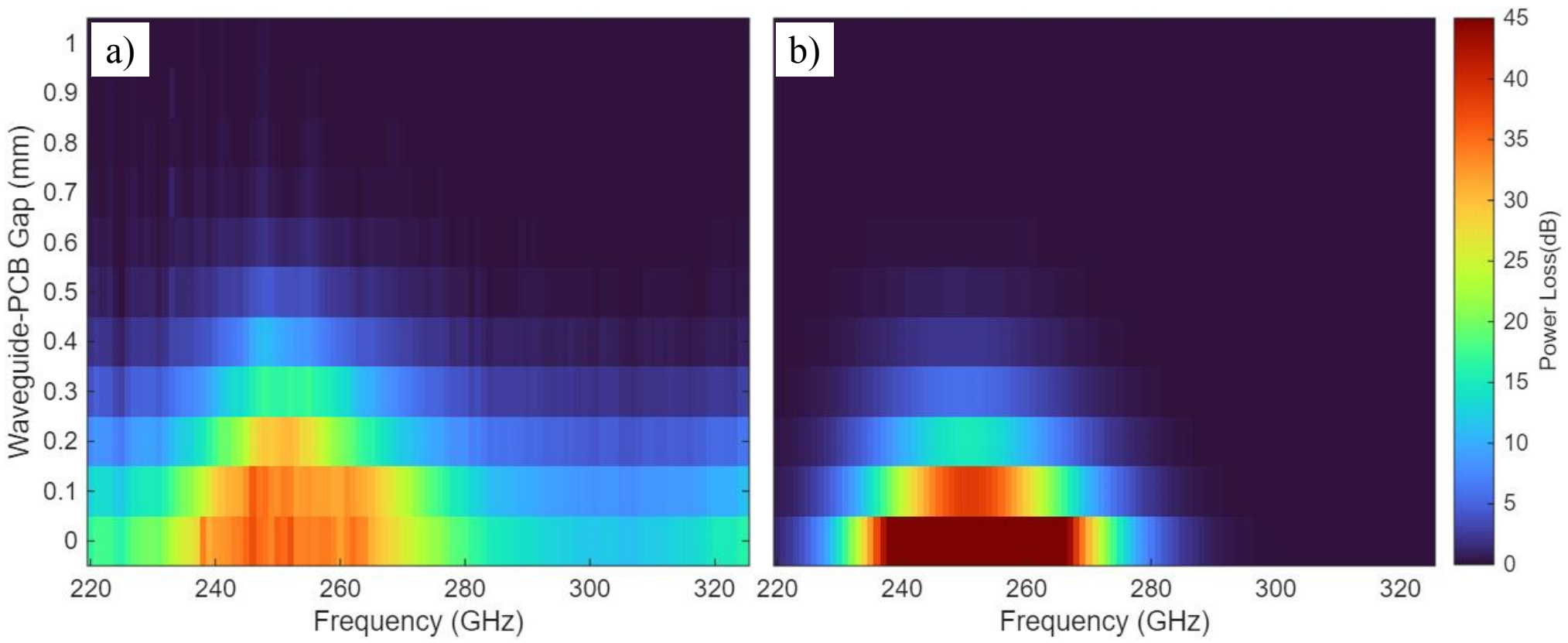

**Fig. 3.** Excess-loss heat maps versus frequency and waveguide-PCB gap for the bare PCB case: (a) measured results; (b) calculated results from the compact engineering descriptor.

Residual discrepancies between Fig. 3(a) and Fig. 3(b) - a broader high-frequency tail and small irregular features - identify physical processes intentionally omitted from the descriptor, as the finite PCB-edge scattering at the 100 mm × 100 mm sample perimeter; the local thickness non-uniformity of the FR4 substrate (≈1% across the sample); the weak higher-order-mode participation at the upper band edge, where Figs. S1 and S3 indicate additional confined branches for the 1×1 $mm^2$ waveguide; the standing-wave residuals from imperfect anti-reflection at the horn-to-waveguide interfaces [20, 49, 50]; and the coupled-mode saturation behavior near the phase-matching peak discussed above, which the present separable form does not capture. The descriptor is bounded in scope to the bare-FR4 case. Its extension to trace-PCB geometries requires explicit treatment of grating-mediated coupling, addressed separately in Section 4.

## 4. Transmission Loss of Periodic Copper-Trace Layouts

In practical PCB implementations, metallization rarely takes the form of an ideal continuous conducting plane. It appears instead as discrete copper traces with layout-dependent width, spacing, and orientation [47], as illustrated in Fig. 1(b). Under these conditions, the proximity-induced transmission behavior of the dielectric waveguide is governed not only by the average copper-coverage ratio but also by the anisotropic electromagnetic response of the trace array and its alignment with the guided-mode evanescent field. To examine these effects, four periodic-trace PCBs are fabricated with a fixed trace width of 1 mm and inter-trace gaps of 1, 2, 3, and 4 mm, corresponding to copper-coverage ratios of approximately 49%, 33%, 25%, and 20%, respectively. For each PCB, three coupling configurations are measured - a perpendicular configuration, in which the waveguide propagation direction is orthogonal to the trace orientation; and two parallel configurations, in which the waveguide propagates along the trace direction and is positioned either above a copper strip (on-trace) or above the dielectric region between adjacent strips (between-traces). The corresponding measured excess-loss spectra are shown in Fig. 4. Because the trace geometries are not invariant both along and across the propagation direction, they lie outside the scope of the simulation validated in Section 3. The spectra below are measured directly, and where a mechanism would require genuinely three-dimensional grating physics - narrowband Wood-type resonances, edge currents, finite trace width, or lateral positioning - the corresponding statements are bounded interpretations of the measured trends, with confirmation left to future work.

### 4.1 Parallel configuration

When the waveguide propagation direction is parallel to the copper traces, the excess loss depends strongly on whether the waveguide is aligned with a metallic strip or with a dielectric gap. In the on-trace case, the evanescent field overlaps directly with a continuous strip extending along the propagation direction. Under the good-conductor boundary condition, the tangential electric field is suppressed at the metal surface while the tangential magnetic field induces longitudinal surface currents on the copper strip [51]. In the present geometry, the strip acts as a localized metallic perturbation that perturbs the guided mode, rather than as a distinct longitudinal energy-transport channel. The associated excess loss arises from conductor dissipation due to induced surface currents, modal perturbation by the metal boundary, and enhanced coupling to substrate-loaded leakage or re-radiation channels in the surrounding PCB environment [51, 52]. This explains why the on-trace condition produces large excess loss when the trace array is dense, particularly in Fig. 4(a), where the 1 mm gap / 49% coverage case yields the highest on-trace attenuation. In dense arrays the evanescent field reaches beyond the strip directly under the waveguide and interacts with adjacent strips, extending the effective metallic loading region.

In the between-traces case, the waveguide is exposed primarily to the dielectric gap and the lossy FR4 substrate beneath it. The dominant interaction is no longer metal-induced current extraction but substrate-assisted leakage through the dielectric window, governed by the same phase-matched leaky-slab-mode mechanism identified in Section 3.4 for the bare-FR4 case. As the inter-trace spacing increases, the effective dielectric opening seen by the evanescent field becomes larger, enhancing field penetration into the substrate and strengthening leakage into the lossy PCB body. This trend is clearly visible in Fig. 4(c) and (d), where the between-traces curves (blue) exceed the on-trace response (black) across a broad frequency range, especially within the 235-265 GHz band where the bare-FR4 leakage channel is most active.

These two opposing tendencies define the parallel-configuration trend. As the trace gap increases from 1 to 4 mm, the on-trace loss decreases because the metallic-loading length and induced-current density imposed by the trace array are reduced, whereas the between-traces loss remains large or grows because the enlarged dielectric windows facilitate substrate-assisted leakage. Under parallel alignment, the observed attenuation therefore reflects a competition between metal-induced current extraction and gap-mediated substrate leakage, and the dominant mechanism shifts as the trace spacing changes.

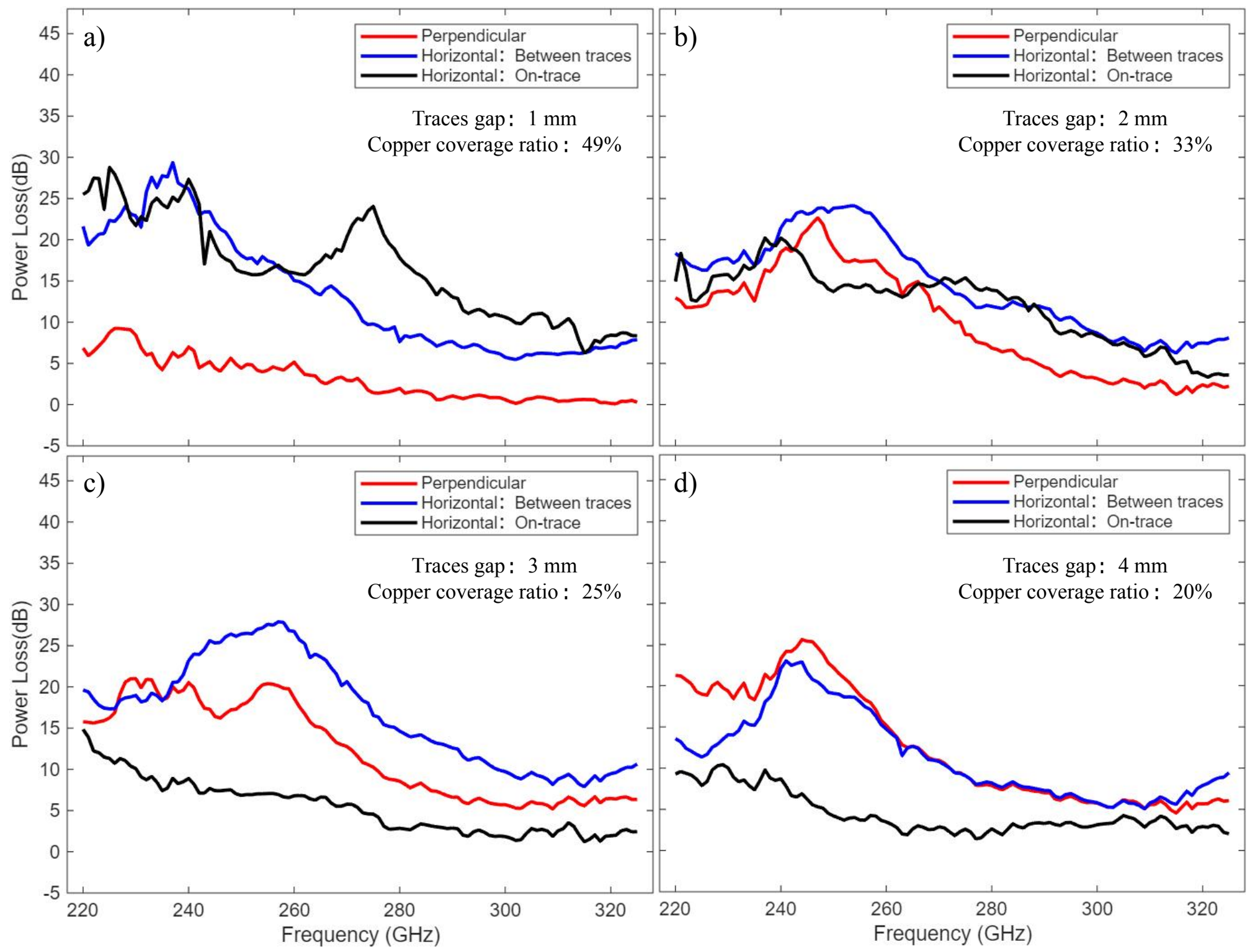

**Fig. 4.** Measured excess power loss for periodic copper-trace PCBs with a 0 mm waveguide-PCB gap under three coupling configurations (perpendicular, between traces, and on-trace): (a) 1 mm trace gap, 49% copper coverage; (b) 2 mm trace gap, 33% copper coverage; (c) 3 mm trace gap, 25% copper coverage; (d) 4 mm trace gap, 20% copper coverage.

### 4.2 Perpendicular configuration

A different behavior is observed when the waveguide propagation direction is perpendicular to the trace orientation. In this configuration, the guided mode crosses an alternating sequence of metallic strips and dielectric gaps along the propagation path, and the waveguide is subject to a periodically modulated metal-dielectric boundary rather than a continuous metallic loading path. Periodic longitudinal perturbations in dielectric waveguides are known to couple guided and radiating waves and to induce radiation loss [53]. Combined with the evanescent-field sensitivity of THz dielectric waveguides to nearby media [20, 27], the excess loss in the perpendicular configuration is expected to depend primarily on the degree of substrate exposure through the dielectric windows. When the trace spacing is small, the metal strips occupy a larger fraction of the region beneath the waveguide, reducing direct substrate exposure of the evanescent field and weakening substrate-assisted leakage. In this regime, the shielding effect of the dense metal traces dominates over substrate exposure, which explains why the perpendicular configuration exhibits the smallest excess loss in the dense-trace case of Fig. 4(a).

TABLE 2

UNIFIED EXCESS-LOSS METRICS FOR DIFFERENT PCB CONFIGURATIONS AT G=0 MM

| PCB type | Alignment | Trace Gap | Copper coverage | $\langle \Delta P_{\mathrm{loss}} \rangle_{220\text{-}325}$ (dB) | $\Delta P_{max}$ (dB) | $\langle \Delta P_{\mathrm{loss}} \rangle_{235\text{-}265}$ (dB) |
|---|---|---|---|---|---|---|
| Bare FR4 | | | 0% | 19.9 | ≥31.9[a] | 28.7 |
| Full Cu | | | 100% | 2.4 | 7.8 | 4.8 |
| Trace PCB | on-trace | 1 mm | 49% | 16.6 | 28.7 | 19.5 |
| | | 2 mm | 33% | 12.2 | 20.2 | 16.2 |
| | | 3 mm | 25% | 5.3 | 14.9 | 7.4 |
| | | 4 mm | 20% | 4.5 | 10.4 | 5.7 |
| Trace PCB | Between traces | 1 mm | 49% | 13.5 | 29.4 | 21.1 |
| | | 2 mm | 33% | 14.6 | 24.1 | 22.1 |
| | | 3 mm | 25% | 16.9 | 27.9 | 24.9 |
| | | 4 mm | 20% | 11.4 | 23.0 | 19.1 |
| Trace PCB | perpendicular | 1 mm | 49% | 3.1 | 9.2 | 4.9 |
| | | 2 mm | 33% | 9.9 | 22.7 | 18.2 |
| | | 3 mm | 25% | 12.2 | 21.0 | 18.7 |
| | | 4 mm | 20% | 12.6 | 25.6 | 21.3 |

[a]The value is limited by the measurement dynamic range and should be interpreted as a lower-bound estimate.

As the inter-trace gap increases and the copper coverage decreases, this shielding effect is progressively weakened. The evanescent field penetrates more efficiently through the dielectric gaps into the substrate, restoring substrate-dominated leakage and increasing the measured attenuation. This trend is directly reflected in Fig. 4(b)-(d), where the perpendicular curves (red) rise substantially as the trace spacing increases from 2 to 4 mm, particularly within the 235-265 GHz region where the underlying waveguide-substrate coupling is strongest.

The perpendicular case is, in principle, also susceptible to grating-mediated effects such as Wood anomalies. For a guided mode of effective index $n_{\mathrm{eff}}$ propagating along a periodic perturbation of period $\Lambda$, the radiative grating-coupling condition for the $m$-th diffractive order, expressed at grazing observation in the air cladding ($n_{\mathrm{clad}} \approx 1$), is $\lambda_{\mathrm{air}} \approx \Lambda \cdot (n_{\mathrm{eff}} \pm n_{\mathrm{clad}} \cdot \sin\theta)/m$. Forward-radiating orders correspond to the minus sign. Backward orders, falling well above the measurement band for all tested periods, are not considered further. For the forward $m = 1$ condition and $n_{\mathrm{eff}} \approx 1.28\text{-}1.37$ extracted from the EIM, the in-band grating-coupling frequencies are $\Lambda = 2$ mm (1 mm trace + 1 mm gap), approximately 405-535 GHz, above the band; $\Lambda = 3$ mm, approximately 270-357 GHz, partially within the band; $\Lambda = 4$ mm, approximately 203-268 GHz, partially within the band; and $\Lambda = 5$ mm, approximately 162-214 GHz, below the band. The $m = 1$ forward condition is therefore satisfiable inside 220-325 GHz for the $\Lambda = 3$ mm and $\Lambda = 4$ mm cases.

The observed perpendicular-configuration excess loss is nevertheless inconsistent with a classical Wood-anomaly attribution on two independent bandwidth grounds. First, Wood anomalies in lossy gratings appear as narrow Fano-like features, typically of sub-GHz to few-GHz spectral width set by the radiation rate of the leaky branch [54]; the broadband 25-30 GHz wide elevations actually observed in Fig. 4(b)-(d) exceed this linewidth by more than an order of magnitude. Second, the same 25-30 GHz wide band, centered between 235 and 265 GHz, recurs across all PCB configurations, including the ungrated bare-FR4 case (Fig. 2(a)), and is therefore not a property of the grating but of the substrate. Both observations - the measured linewidth and the recurrence of the band in the ungrated bare-FR4 case (Fig. 2(a)) - are drawn directly from the spectra, so the conclusion that the broadband perpendicular-configuration loss is dominated by the same substrate-assisted phase-matched leakage channel of Section 3.4, modulated by the orientation-dependent dielectric exposure through the inter-trace windows, does not depend on any full-wave simulation of the trace array. Any genuine narrowband Wood-anomaly contribution at the $\Lambda$ = 3 mm or $\Lambda$ = 4 mm frequencies (270 or 268 GHz) would appear superposed on this substrate-dominated continuum; the measured spectra do not resolve features at the corresponding linewidth, but the present dataset is also not designed to. Confirming or excluding a residual narrowband contribution at the $\Lambda$ = 3 mm or $\Lambda$ = 4 mm frequencies would require a targeted sub-GHz swept-frequency measurement on those samples together with a dedicated three-dimensional full-wave study of a single grating period. Such refinement lies beyond the present scope and does not affect the substrate-dominated mechanism established here [54, 55].

**4.3 Unified excess-loss metrics and design implications**

To provide a compact quantitative comparison among the configurations, three unified excess-loss metrics are extracted from the measured spectra and summarized in Table 2. The band-averaged excess loss $\langle \Delta P_{\text{loss}} \rangle_{220\text{-}325}$ over the full measured band, the maximum excess loss $\Delta P_{\text{max}}$ within 220-325 GHz, and the band-averaged loss $\langle \Delta P_{\text{loss}} \rangle_{235\text{-}265}$ in the strong-coupling region identified in Section 3.4.

For the bare FR4 case, direct contact produces $\langle \Delta P_{\text{loss}} \rangle_{220\text{-}325}$ = 19.9 dB and $\langle \Delta P_{\text{loss}} \rangle_{235\text{-}265}$ = 28.7 dB, while the full-Cu case yields 2.4 dB and 4.8 dB respectively. For the periodic copper-trace PCBs, the loss depends strongly on both copper coverage and trace alignment. In the on-trace configuration, all three metrics decrease monotonically as the trace gap increases from 1 to 4 mm, with $\langle \Delta P_{\text{loss}} \rangle_{220\text{-}325}$ falling from 16.6 to 4.5 dB and $\Delta P_{\text{max}}$ from 28.7 to 10.4 dB. In the between-traces configuration, the loss remains large for all trace gaps, especially in the 235-265 GHz band where $\langle \Delta P_{\text{loss}} \rangle_{235\text{-}265}$ ranges from 19.1 to 24.9 dB. Under perpendicular alignment, the dense-trace case (49% coverage) produces the lowest loss among the trace-PCB measurements, with $\langle \Delta P_{\text{loss}} \rangle_{220\text{-}325}$ = 3.1 dB and $\langle \Delta P_{\text{loss}} \rangle_{235\text{-}265}$ = 4.9 dB; as the trace gap increases and copper coverage decreases, the loss rises substantially, reaching $\langle \Delta P_{\text{loss}} \rangle_{220\text{-}325}$ = 12.6 dB and $\langle \Delta P_{\text{loss}} \rangle_{235\text{-}265}$ = 21.3 dB at 4 mm spacing.

To quantify the shielding effectiveness of a given PCB configuration, we extend the band-averaged shielding factor of Section 3.4 to arbitrary metallization geometry,

$$\eta_{\text{shield}}^{config} = 1 - \frac{\langle \Delta P_{loss} \rangle_{config}}{\langle \Delta P_{loss} \rangle_{bare-FR4}} \tag{9}$$

Using $\langle \Delta P_{\text{loss}} \rangle_{220\text{-}325}$, the full-Cu boundary suppresses approximately 88% of the bare-FR4-induced excess loss, the dense-trace perpendicular configuration achieves $\eta_{\text{shield}} \approx 0.84$, and the sparse-trace perpendicular configuration drops to $\eta_{\text{shield}} \approx 0.37$. This confirms that continuous and dense-transverse metallization act primarily as waveguide-facing shielding boundaries rather than dominant loss channels in the tested geometry.

The packaging-level design implications of these unified metrics are summarized in Table 3. The recommendations apply within the tested geometry - a 1×1 $mm^2$ air-cladded PP waveguide with 100 mm interaction length over 220-325 GHz - and are not intended as universal rules across all THz packaging scenarios. They nonetheless provide actionable guidance for avoiding the dominant evanescent-leakage pathways identified in Section 3 when routing dielectric waveguides in PCB-proximal environments.

TABLE 3

DESIGN IMPLICATIONS FOR PCB-PROXIMAL THZ DIELECTRIC-WAVEGUIDE ROUTING

| Condition | Observed effect | Design implication | Recommended use |
| --- | --- | --- | --- |
| Bare FR4, g=0 mm | Strong frequency-selective loss; high-loss band around 235–265 GHz | Avoid direct contact with lossy FR4 | Not recommended |
| Bare FR4, g=0.5 mm | Loss strongly reduced compared with contact | Clearance is effective for suppressing leakage | Useful clearance when package height is constrained |
| Bare FR4, g=1 mm | Excess loss approaches low level over most of the band | Increasing clearance suppresses evanescent leakage | Preferred when package height allows |
| Full Cu facing waveguide | Much lower loss than bare FR4 in contact | Continuous metal layer can serve as shielding boundary | Recommended as a waveguide-facing shield in the tested band and geometry |
| Dense trace PCB, perpendicular | Low loss for 49% coverage | Dense transverse traces can approximate shielding | Acceptable when continuous Cu is unavailable |
| Sparse trace PCB, perpendicular | Increased loss as trace gap increases | Large dielectric openings restore FR4 leakage | Avoid sparse transverse traces near waveguide |
| Trace PCB, on-trace | Loss decreases as coverage decreases | Longitudinal metal loading can perturb the mode | Avoid dense longitudinal traces directly below waveguide |
| Trace PCB, between-traces | Loss remains high | Dielectric windows expose the waveguide to FR4 | Avoid routing waveguide above trace gaps |

## 5. Conclusion

Bare dielectric waveguides are increasingly attractive for short-reach THz interconnects, but their air-clad evanescent fields couple into nearby PCB materials, and this proximity-induced loss has not been systematically quantified. This article characterized, experimentally and numerically, the near-field-coupling-induced transmission behavior of 3D-printed PP dielectric waveguides routed near PCB substrates over 220-325 GHz. Direct contact with bare FR4 produces a 235-265 GHz high-loss band, most consistently explained as phase-matched leakage from the PP-guided fundamental mode into the $m = 3$ transverse slab branch of the FR4. A continuous waveguide-facing copper layer suppresses about 88% of the band-averaged excess loss, with a residual peaking at $\leq 8$ dB from ohmic dissipation. A 1 mm clearance reduces the loss to noise-floor levels, consistent with the EIM air-side decay length of 152-253 μm. The impairment is essentially attenuative, with the differential GVD bounded within $\pm 1\times10^{-7}$ ps$^2$/m. Periodic copper traces show strongly orientation-dependent loss - perpendicular dense traces approximate a shielding boundary, whereas sparse arrays expose the substrate, and the 25-30 GHz band recurs in all configurations including the ungrated case, ruling out classical Wood anomalies. This work provides layout guidance for routing bare THz dielectric waveguides in compact PCB-proximal packages, thereby reducing a key uncertainty in package-level THz interconnect integration while leaving residual narrowband grating contributions, three-dimensional PCB-edge effects, and alternative polymer – substrate pairings as open topics for future work.

**Acknowledgment**

This work was supported in part by the National Natural Science Foundation of China under Grant (62471033), and the Special Program Project for Original Basic Interdisciplinary Innovation under the Science and Technology Innovation Plan of Beijing Institute of Technology under Grant (2025CX11010).